\documentclass[preprint2]{aastex}
\usepackage{amsmath}
\usepackage{natbib}
\usepackage{enumerate}
\usepackage{color}
\begin{document}

\title{Evidence for DCO$^{+}$ as a probe of ionization in the warm disk surface}

\author{C\'ecile Favre, Edwin~A. Bergin, L.~Ilsedore Cleeves}
\affil{Department of Astronomy, University of Michigan, 500 Church St., Ann Arbor, MI 48109}
\email{cfavre@umich.edu}
\author{Franck Hersant}
\affil{Univ. Bordeaux, LAB, UMR 5804, F-33270, Floirac, France}
\affil{CNRS, LAB, UMR 5804, F-33270, Floirac, France}
\author{Chunhua Qi,}
\affil{Harvard-Smithsonian Center for Astrophysics, 60 Garden Street, Cambridge, MA 02138}
\and
\author{Yuri Aikawa}
\affil{Department of Earth and Planetary Sciences, Kobe University, Kobe 657--8501, Japan}

%
\begin{abstract}
In this Letter we model the chemistry of DCO$^{+}$ in protoplanetary disks. We find that the overall distribution of the DCO$^{+}$ abundance is qualitatively similar to that of CO but is dominated by thin layer located at the inner disk surface. To understand its distribution, we investigate the different  key gas-phase deuteration pathways that can lead to the formation of DCO$^{+}$. Our analysis shows that the recent update in the exothermicity of the reaction involving CH$_2$D$^{+}$ as a parent molecule of DCO$^{+}$ favors deuterium fractionation in warmer conditions. As a result the formation of DCO$^{+}$ is enhanced in the inner warm surface layers of the disk where X--ray ionization occurs. Our analysis points out that DCO$^{+}$ is not a reliable tracer of the CO snow line as previously suggested. We thus predict that DCO$^{+}$ is a tracer of active deuterium and in particular X--ray ionization of the inner disk. 
\end{abstract}

\keywords{protoplanetary disks --- astrochemistry --- ISM: abundances --- stars: formation}

\maketitle

%
\section{Introduction}
\label{sec:int}

The D/H ratio is a key tracer of history and origin of chemical and physical processes that occurred in the early phases of protoplanetary disks, in which planets may form. In that light, \citet{van-Dishoeck:2003} have suggested that deuterated chemistry pathways are active during the early phase of planet formation which may alter the deuterium enrichment supplied by the earlier, pre-stellar stages. These deuterium enhancements may be provided to ice-rich planetesimals during formation. Thus understanding deuterium fractionation is key to unravelling the clues of the D/H ratios in the various bodies found in our solar system \citep[e.g.][]{Mumma:2011}.

In the physical conditions (temperature, density, ionization) prevailing in circumstellar disks surrounding T Tauri stars, deuterium fractionation is expected to be dominated by ion-molecule reactions \citep[see e.g.][]{Aikawa:1999}. Among deuterated ions, DCO$^+$ has been detected in various disk environment such as toward the protoplanetary disk of DM~Tau \citep{Guilloteau:2006,Oberg:2010a,Oberg:2011b,Teague:2015}, the disk around the classical T~Tauri star TW~Hya  \citep{van-Dishoeck:2003,Qi:2008,Oberg:2012} as well as in the disk surrounding HD~163296 \citep{Mathews:2013} but also in the following other disks, LkCa~15, IM~Lup, AS ~09, and V4046~Sgr \citep[SMA DISCS project, see][]{Oberg:2010a,Oberg:2011a}.

The key gas-phase deuterium pathways that are active at cold temperature (typically T$<$50K) are the following reactions \citep{Watson:1976,Wootten:1987}: 
\begin{equation}
\label{eq1}
\textnormal{H}_3^{+}+ \textnormal{HD} \leftrightarrows \textnormal{H}_2\textnormal{D}^{+}+ \textnormal{H}_2+\textnormal{232K}
\end{equation}
 and then,
\begin{equation}
\label{eq2}
\textnormal{H}_2\textnormal{D}^{+}+ \textnormal{CO} \rightarrow \textnormal{DCO}^{+} + \textnormal{H}_2
\end{equation}

Reaction \ref{eq1} results in H$_2$D$^{+}$ enhancement as it is driven mainly forward (i.e. left-to-right) at low temperature \citep{Pagani:1992}. Then by reacting with CO, more specifically by transferring a D-atom to CO, the H$_2$D$^{+}$ ion can lead to the formation of DCO$^{+}$ (reaction.~\ref{eq2}). Due to the low temperature condition that is favorable to reactions~\ref{eq1} and \ref{eq2}, \citet{Mathews:2013} have suggested that DCO$^{+}$ is a good tracer of the disk midplane near the CO snow--line. The latter represents a line dividing regions where the disk is warm enough for CO to be in the gas phase and regions where the temperatures are low enough for CO to be depleted on the grains.

However, modeling the chemical structure of a protoplanetary disk is not trivial as the chemistry depends of the physical environment \citep[e.g.][]{Dutrey:2014,Aikawa:2002} ; and DCO$^{+}$ may form independently of H$_2$D$^{+}$ though reactions with CH$_{2}$D$^{+}$ under warmer conditions \citep[typically T$>$ 50K, see e.g.][]{Wootten:1987}. In particular, in warm conditions the CH$_{2}$D$^{+}$ ion is mainly formed by
\begin{equation}
\label{eq3}
\textnormal{CH}_3^{+} + \textnormal{HD} \leftrightarrows \textnormal{CH}_2\textnormal{D}^{+} + \textnormal{H}_2 + \textnormal{$\Delta$E}.
\end{equation}
Reactions involving CH$_{2}$D$^{+}$ or one of its products, such as reaction
\begin{equation}
\label{eq4}
\textnormal{CH}_2\textnormal{D}^{+} + \textnormal{O} \rightarrow \textnormal{DCO}^{+} + \textnormal{H}_2
\end{equation}
and
\begin{equation}
\label{eq5}
\textnormal{CH}_4\textnormal{D}^{+} + \textnormal{CO} \rightarrow \textnormal{DCO}^{+} + \textnormal{CH}_4,
\end{equation}
can significantly contribute to the formation of DCO$^{+}$.

Until now, the zero point vibrational energy difference between the products and reactions, $\Delta$E, of reaction~\ref{eq3} was estimated to be 370~K \citep{Smith:1982}. However, \citet{Roueff:2013} have recently revised the energy budget of reaction~\ref{eq3} and estimate an exothermicity $\Delta$E of 654~K, which is about a factor 2 larger than the previous estimate. This finding implies that CH$_2$D$^{+}$ can become enhanced even at temperatures as high as 300~K. As a consequence, the DCO$^{+}$ formation rate could also be increased. 
Incidentally, using Plateau de Bure Interferometer observations of the DM~Tau disk, \citet{Teague:2015} have recently pointed out that DCO$^{+}$  can be used as a potential tool to diagnostic ionization in disks. This leads one to ask how the change of exothermicity affects the deuterium chemistry and hence, whether DCO$^{+}$ is a reliable tracer of ionization in disk or, of the CO snow line as previously suggested by \citet{Mathews:2013}. 

In this Letter, we discuss the distribution of the DCO$^{+}$ ion in disk according the new assumptions set by \citet{Roueff:2013}. In Section~\ref{sec:mod} we describe the physical and chemical model used in this study. Results are given in Section~\ref{sec:res} and their implications are discussed in Section~\ref{sec:conc}.

%
\section{Modeling}
\label{sec:mod}
We use the (1+1)--dimensional disk model of \citet{Fogel:2011} to investigate
the main formation pathways of DCO$^{+}$. The physical structure and
reaction network are described below. 
\subsection{Disk Structure}
In this study, a two--dimensional and azimuthally symmetric disk physical
structure that represents a T-Tauri disk has been adopted
\citep[i.e. structure of the TW Hya T-Tauri system see][and references therein]{Cleeves:2013,Schwarz:2014}. Being the same settled disk structure
as the one adopted by
\citet{Schwarz:2014}, we simply refer to this paper for a complete description
of the density/temperature structure and grain distribution.

\subsection{Chemical reaction network} 
We use the same deuterated chemical network as in \citet{Aikawa:2012}, who worked on star forming cores, except
for the following updates and modifications needed for disk chemistry: \textit{i)} the energy barriers of the endothermic
reactions involving the CH$_2$D$^{+}$ ion are updated following
\citet{Roueff:2013}. \textit{ii)} The X-ray ionization of H$_2$ and He along
with UV photolysis induced by X-rays are included \citep{Fogel:2011}.
\textit{iii)} Ly$\alpha$ photodesorption radiation on dust grain is taken
into account \citep[see][for further details]{Fogel:2011,Bethell:2011a}. \textit{iv)} CO, H$_2$, HD and D$_2$ self-shielding are included in a time-dependent way
\citep[see][]{Fogel:2011}. \textit{v)} Finally, grain surface reactions are
limited to the formation of H$_2$, OH and H$_2$O since gas-phase chemistry
prevails in the DCO$^{+}$ formation.

The network contains in total 1565 species and 43353 reactions\footnote{To
process faster the data GNU parallel \citep{Tange:2011} was used along with our
code.}.  Among the reactions listed above, the model also includes grain
surface recombination reactions, adsorption, photodesorption, two body
reactions (including ion, molecule, negative ion, neutral), cosmic--ray
ionization ($\zeta$=1.3$\times$10$^{-17}$~s$^{-1}$), cosmic-ray induced
photo-dissociation of ice, photolysis, thermal desorption, non-thermal
desorption (i.e cosmic-ray evaporation), gas phase and grain surface photodissociation
\citep[see][for further
details]{Hasegawa:1993,Garrod:2008,Fogel:2011,Aikawa:2012}. Incidentally, in
our analysis, following \citet{Garrod:2006}, we assume a CO binding energy set to 1150~K \citep[which
corresponds to a desorption temperature of about $\sim$ 21--25K, see
Supplementary Materials of][]{Cleeves:2014a}. Therefore, the CO snow-line from
our model is shifted by several AU in comparison with \citet{Schwarz:2014}.
Although, the chemical evolution of the disk is run independently (i.e. 1--D
vertical model), the code was run for a given range of radii resulting in a
pseudo two-dimensional result as described by \citet{Schwarz:2014} and
\citet{Fogel:2011}. The gas-grain chemical model is run for
3$\times$10$^{6}$~years (typical disk age) with 45 vertical zones and 22 radial zones.

%
\section{Results}
\label{sec:res}

\subsection{Disk chemical abundances}
\begin{figure*}
\centering
\includegraphics[angle=0,width=8.2cm]{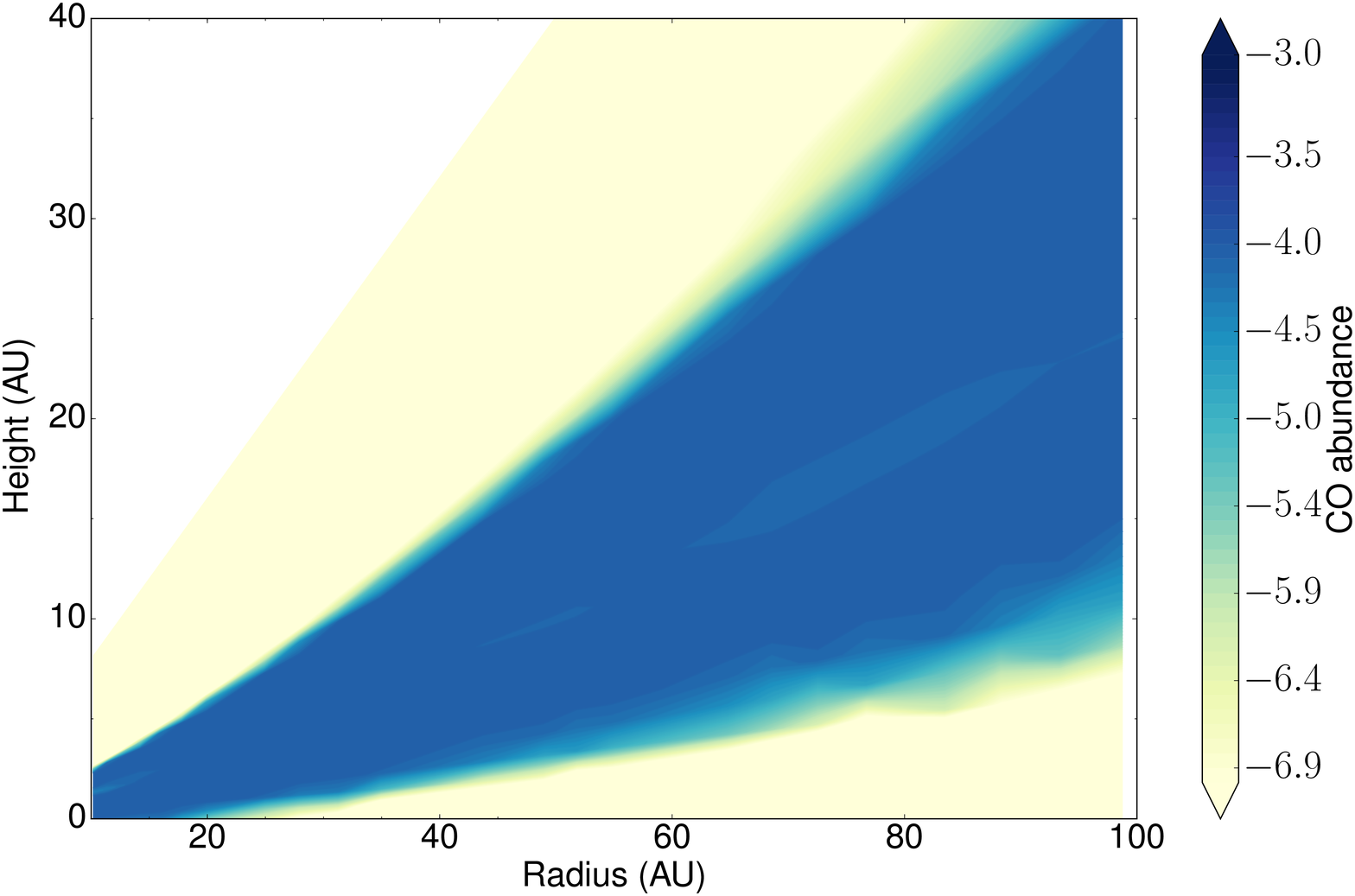}
\includegraphics[angle=0,width=8.2cm]{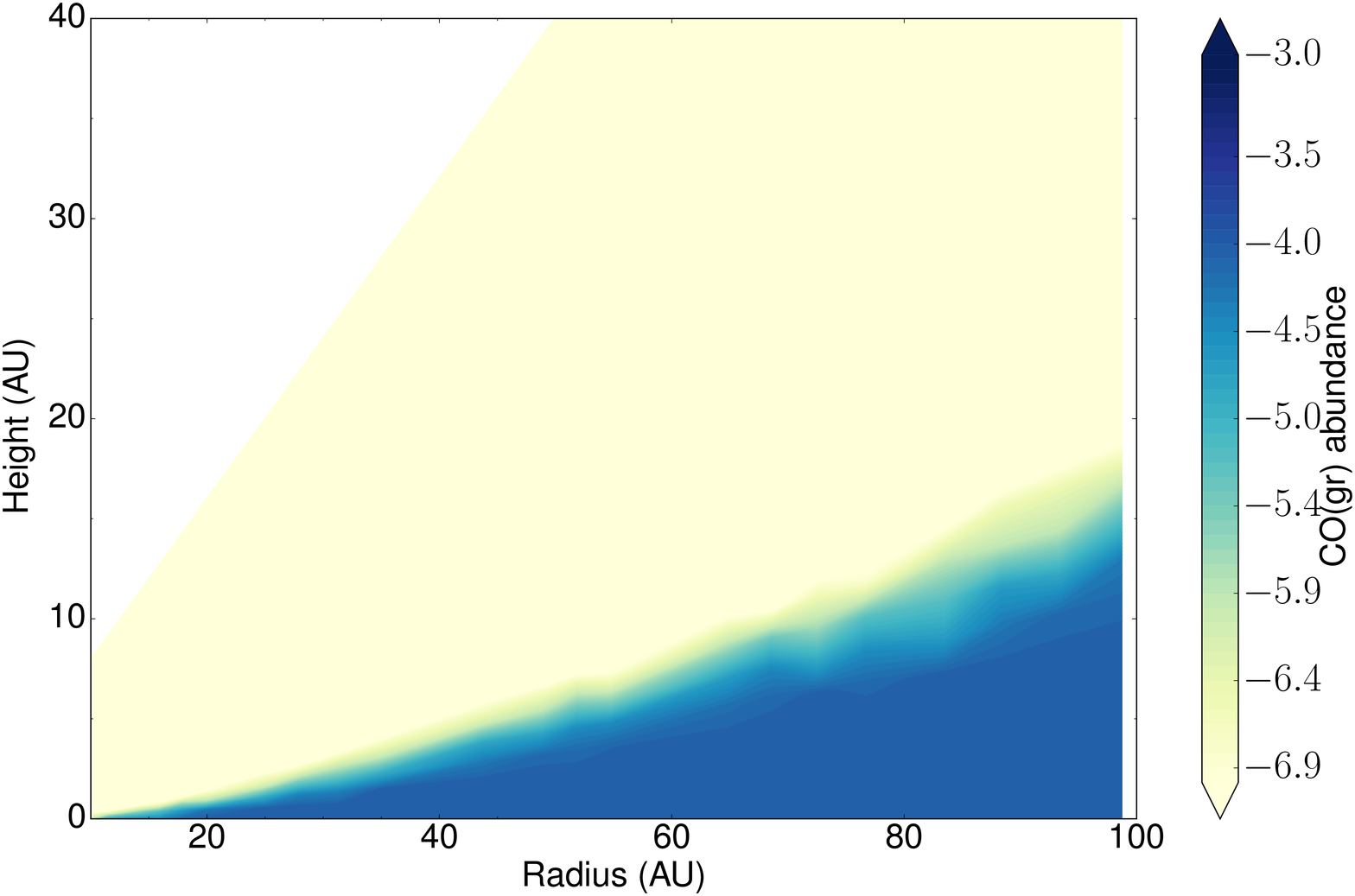}
\includegraphics[angle=0,width=8.2cm]{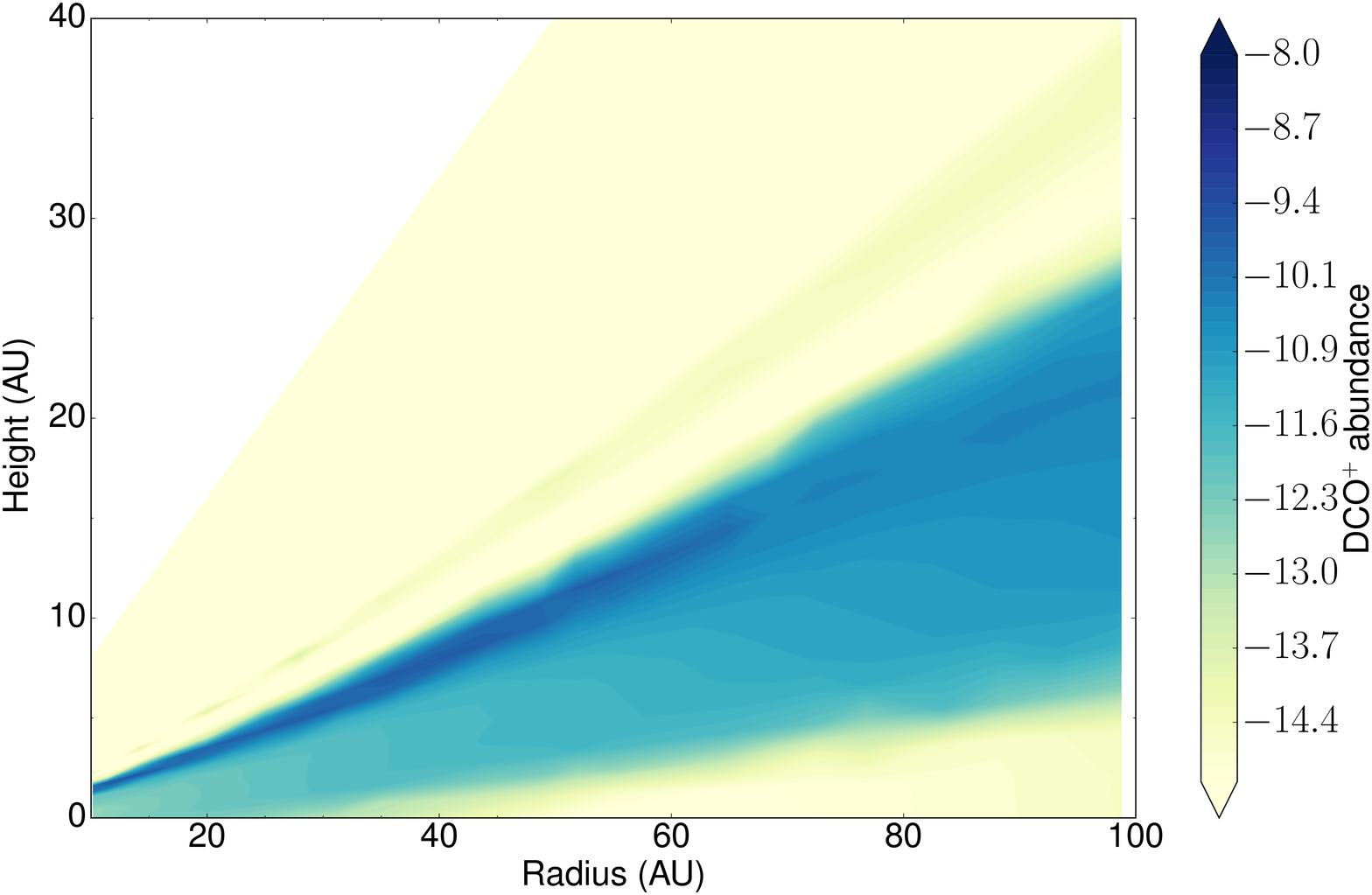}
\includegraphics[angle=0,width=8.2cm]{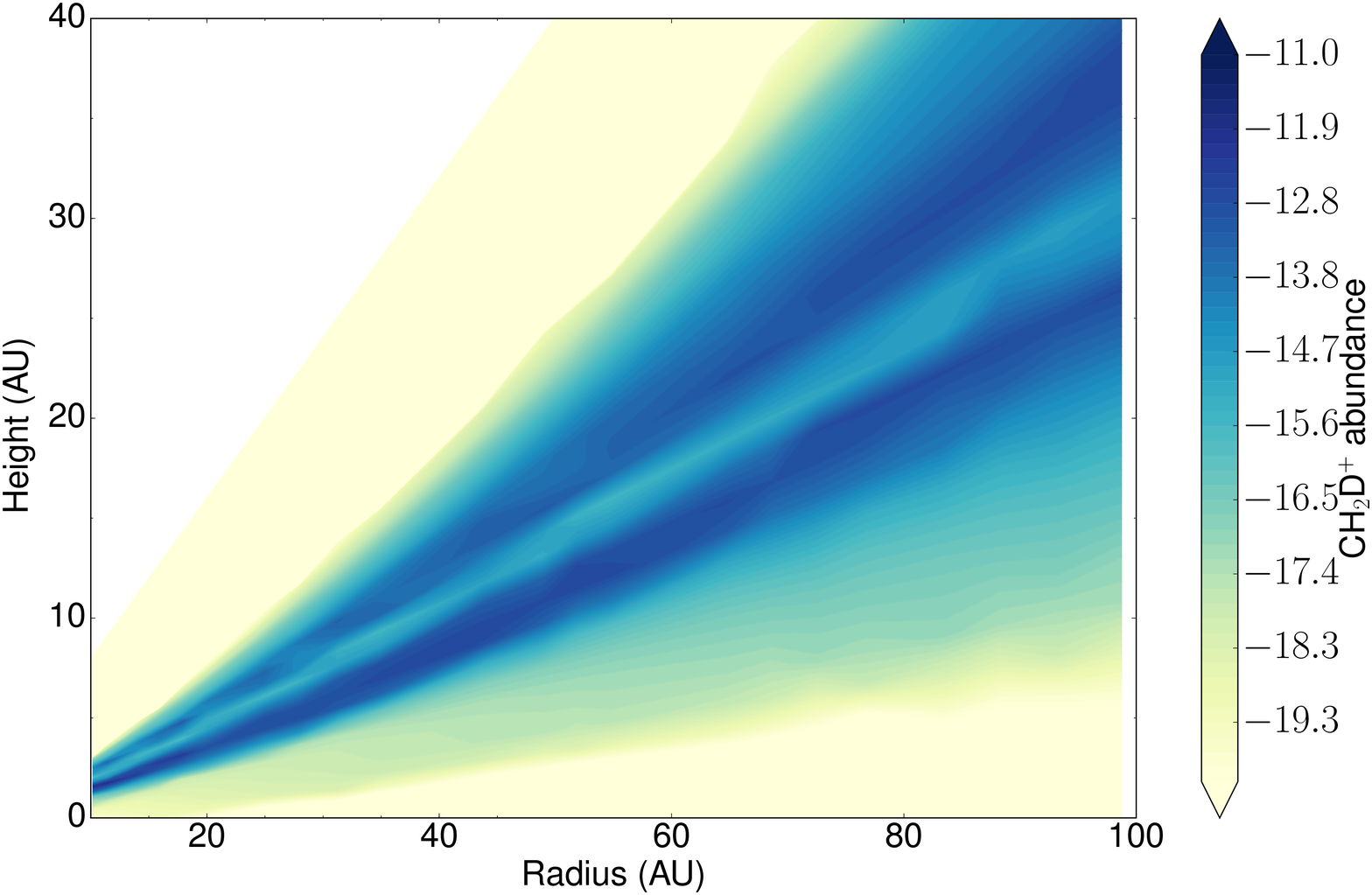}
\includegraphics[angle=0,width=8.2cm]{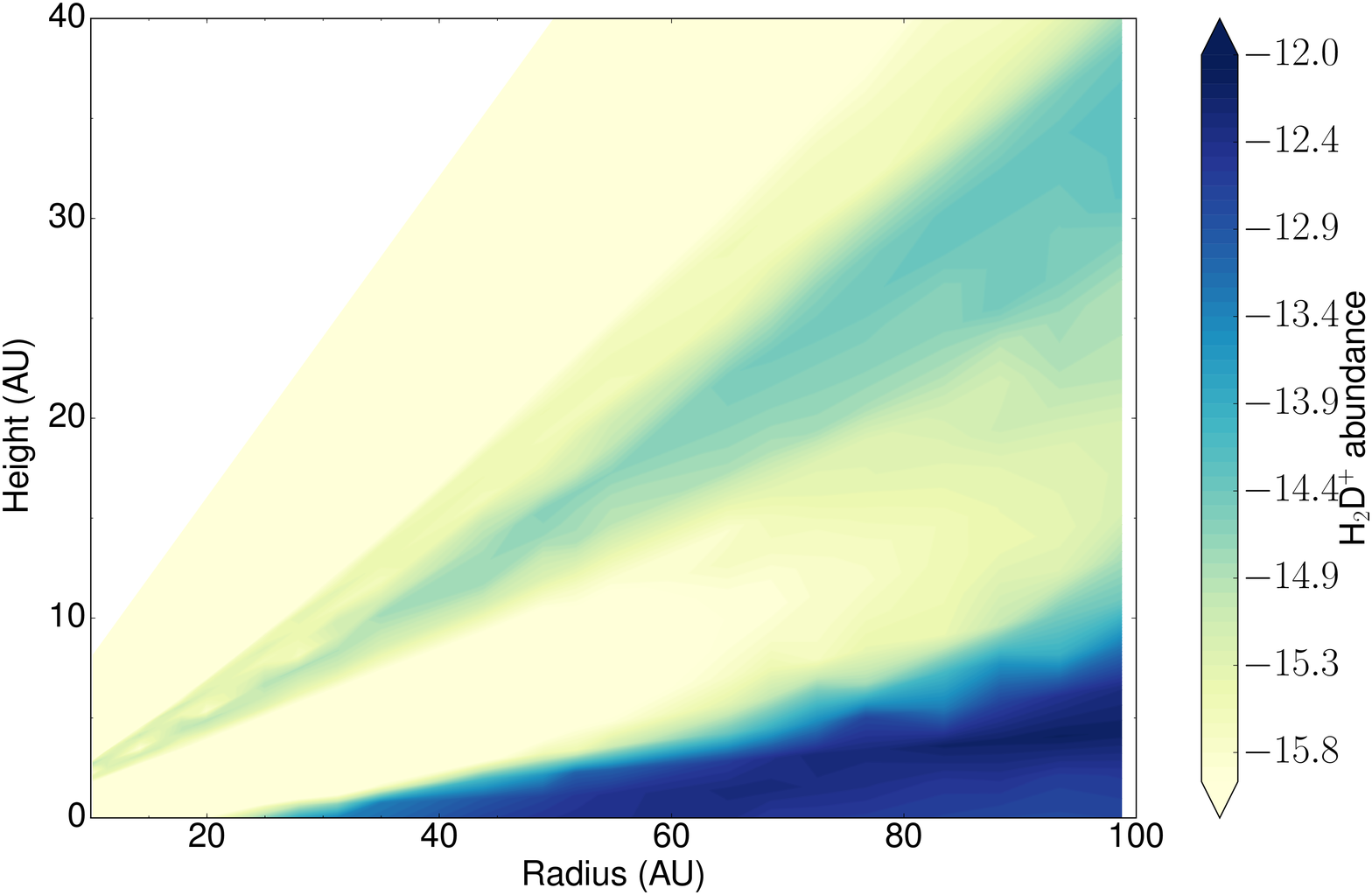}
\includegraphics[angle=0,width=8.2cm]{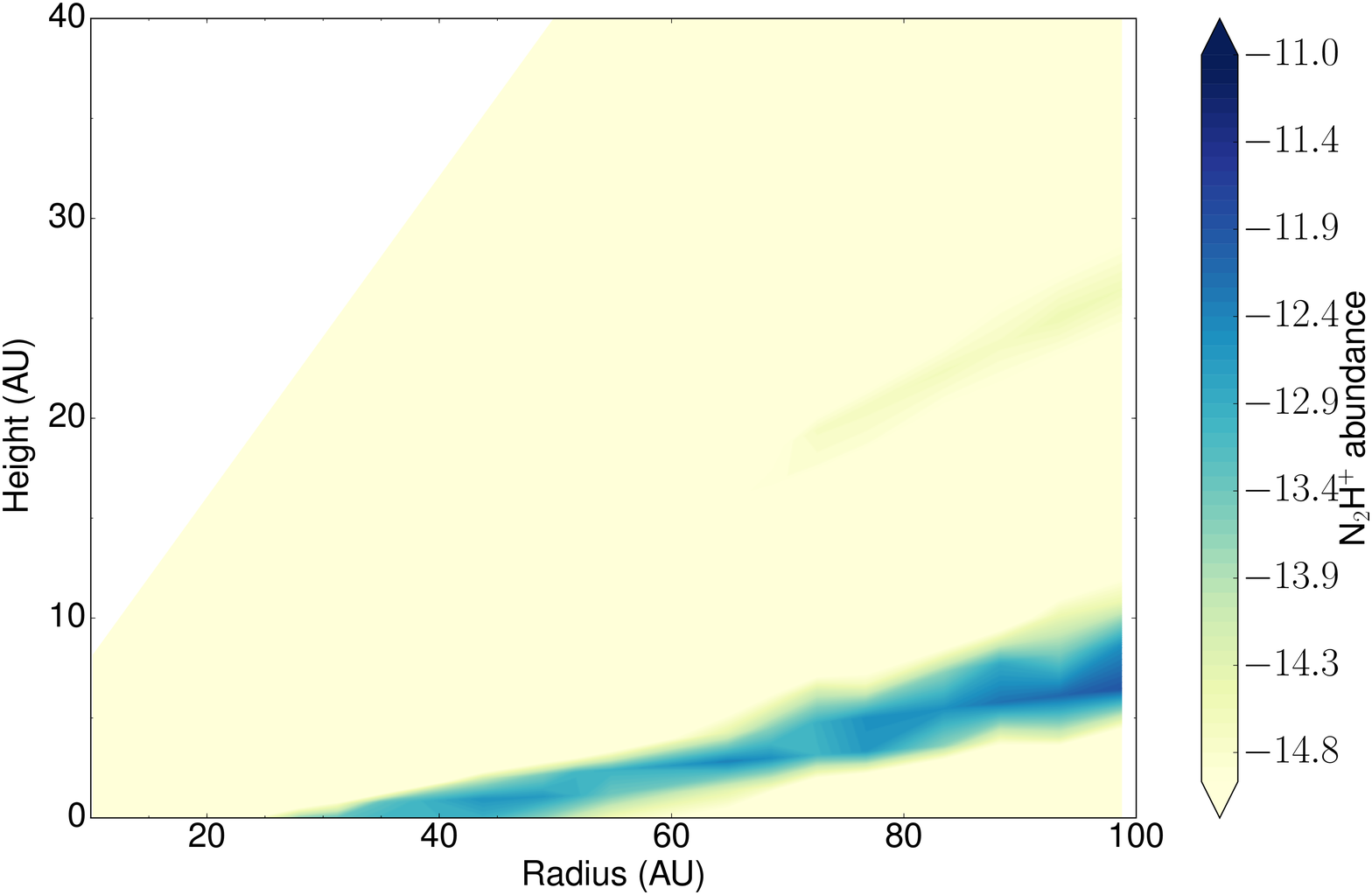}
\caption{Relative abundance to molecular hydrogen. {\em From top left to bottom right:} CO, CO$\rm_{ice}$, DCO$^{+}$, CH$_2$D$^{+}$, H$_2$D$^{+}$ and N$_2$H$^{+}$.\label{fg1}}
\end{figure*}

In Figure~\ref{fg1}, we illustrate where the CO snow line is in our model  by
showing the relative abundance to molecular hydrogen for CO, CO$\rm_{ice}$ and
N$_2$H$^{+}$. The latter ion has been reported to be a good tracer of the CO
snow line since it is destroyed by gas phase CO through reaction~\ref{eq6}
\citep[i.e.][]{Hersant:2009,Qi:2013a,Schwarz:2014}. 
\begin{equation}
\label{eq6}
\textnormal{N}_2\textnormal{H}^{+} + \textnormal{CO} \rightarrow \textnormal{HCO}^{+} + \textnormal{N}_2
\end{equation}
Our model is consistent with this interpretation.

In Figure~\ref{fg1}, we also show the relative abundance to molecular hydrogen
for DCO$^{+}$ along with the one of its 2 potential key precursors,
CH$_2$D$^{+}$ and H$_2$D$^{+}$ (see Reactions~~\ref{eq2} and~\ref{eq3}). It is
immediately apparent that H$_2$D$^{+}$ probes the CO snow line as expected from
reaction~\ref{eq2}. Indeed, the H$_2$D$^{+}$ distribution is clearly
anticorrelated with the one of CO since H$_2$D$^{+}$ is located beyond the CO
snow line (i.e where CO is adsorbed on ice grain mantles) and in the upper disk
layers where the CO is photodissociated. As for CH$_2$D$^{+}$, it is more
abundant in the upper disk layers where ionization supplied by X-rays is higher. Regarding the
overall distribution of the DCO$^{+}$ abundance, it is more or less correlated
with that of CO, with one important exception: there is a greater concentration in the surface of the inner disk. More specifically, the DCO$^{+}$ distribution is not
restricted to the CO snow line but rather spreads up to the disk surface with a
maximal abundance reached in the inner warmer (T$\sim$ 50-100K) surface layers (see
Fig.~\ref{fg1}). 

\subsection{On the DCO$^{+}$ formation pathways}
\begin{figure*}
\centering
\includegraphics[angle=0,width=12cm]{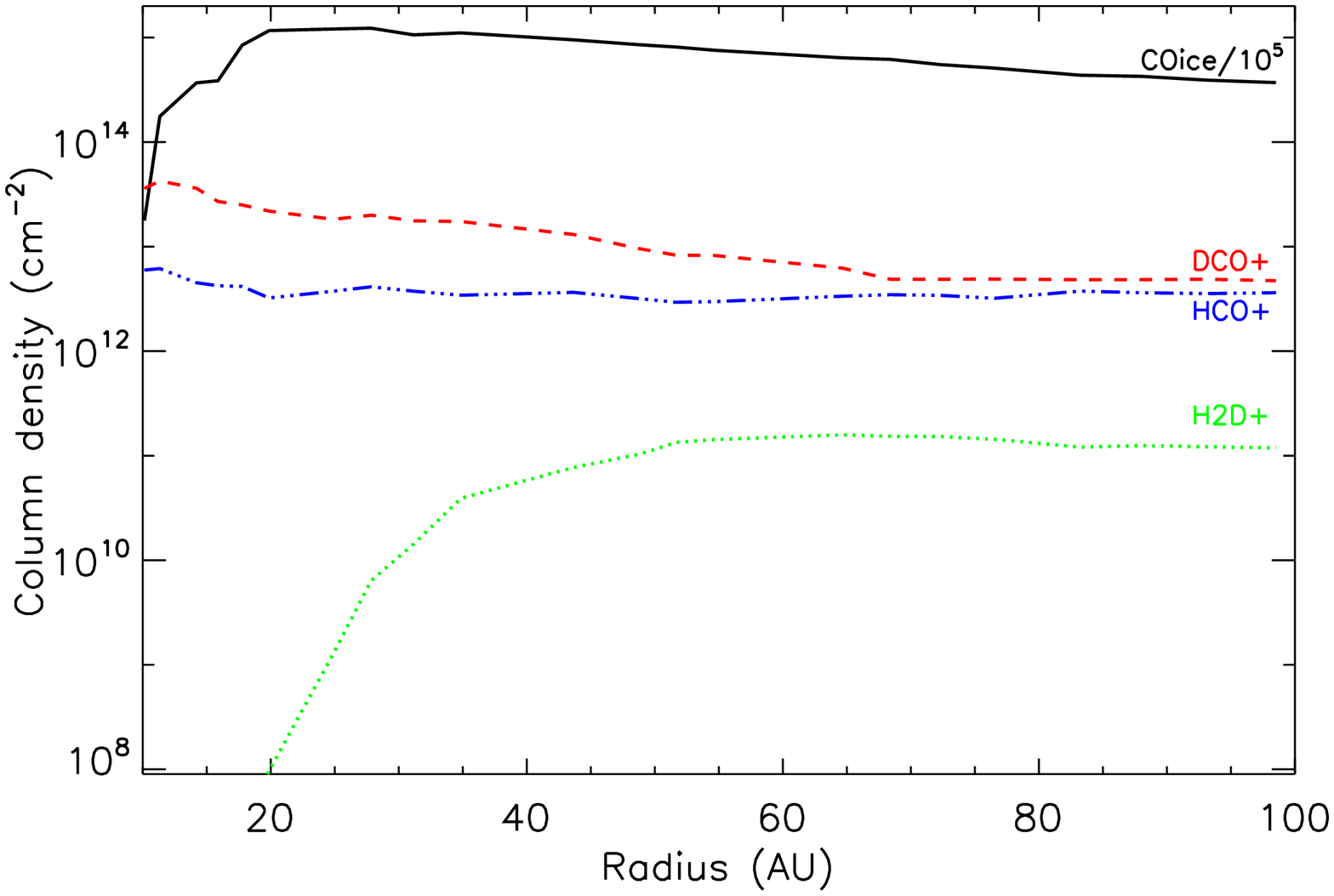}

\includegraphics[angle=0,width=12cm]{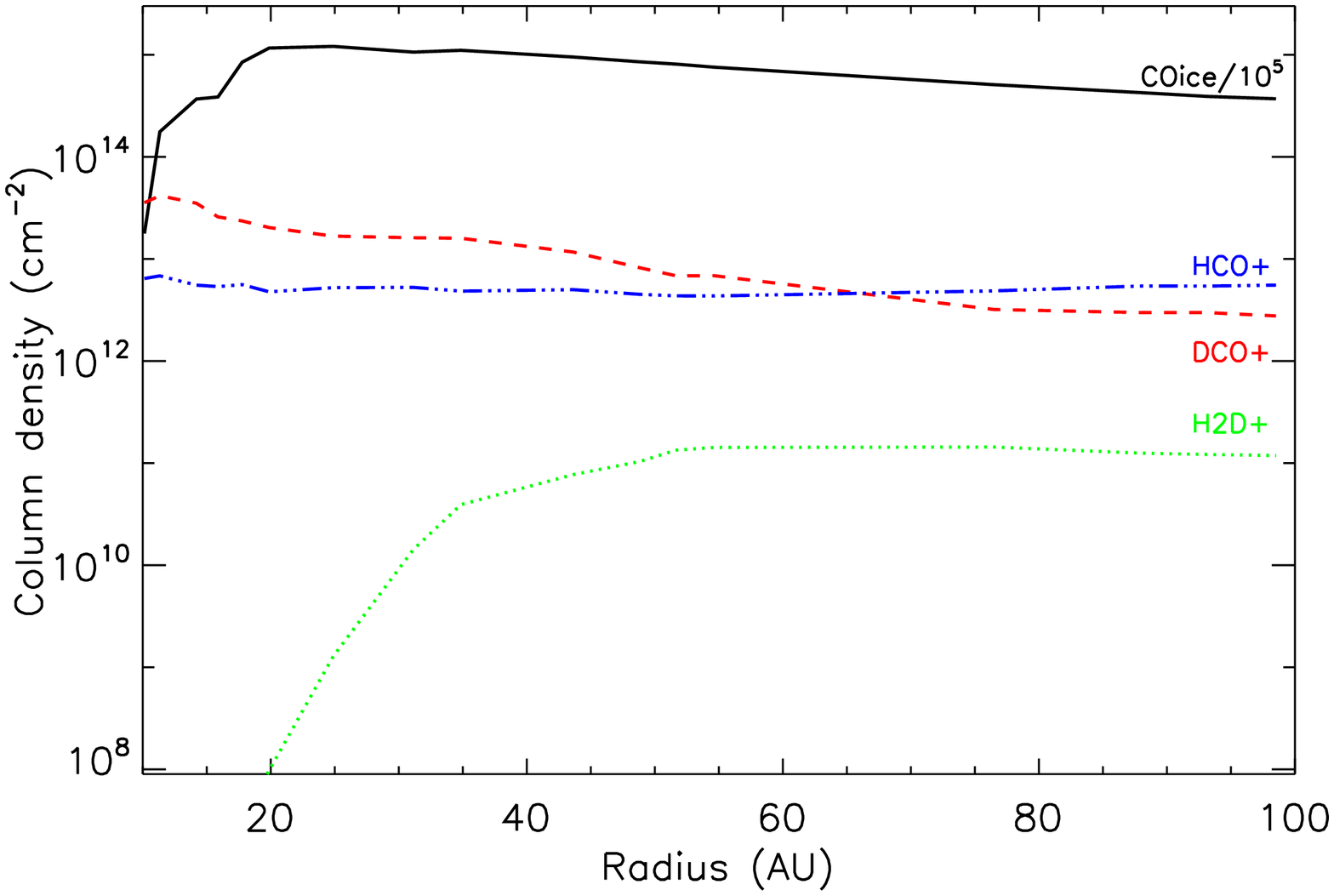}
\caption{Radial distribution of molecular column densities with the H$_2$D$^{+}$ channel (reaction~\ref{eq2}, top panel) and without the H$_2$D$^{+}$ channel (bottom panel).{\em Top panel:} The CO$\rm_{ice}$, DCO$^{+}$, HCO$^{+}$ and H$_{2}$D$^{+}$ distribution are indicated in dark solid lines, red dashed lines, blue dash dot dot lines and green dotted lines, respectively. Note that in both panels the CO$\rm_{ice}$ column density is divided by a factor 10$^{5}$. \label{fg2}}
\end{figure*}

Reaction~\ref{eq2}, involving H$_{2}$D$^{+}$, is usually considered as the main
source of DCO$^{+}$ in disks at low temperatures
\citep[see][]{Mathews:2013}. If this is the case, the qualitative distribution
of DCO$^{+}$ abundance should resemble that of CO. However, as shown in
Figure~\ref{fg1}, the bulk of the DCO$^{+}$ abundance is located in a narrow
layer near the surface. This finding suggests that reaction~\ref{eq2} is not
the only formation route for  DCO$^{+}$ in disks. This is further illustrated
in Figure~\ref{fg2} that shows the DCO$^{+}$,  HCO$^{+}$, CO$_{ice}$  and
H$_{2}$D$^{+}$ column densities as a function of the radius when
reaction~\ref{eq2} is activated (top panel) and turned off (bottom
panel)\footnote{When switching off reaction~\ref{eq2}, we keep reactions
$\textnormal{H}_3^+ + \textnormal{CO} \rightarrow \textnormal{HCO}^+ +
\textnormal{H}_2$ and $\textnormal{H}_2\textnormal{D}^+ + \textnormal{CO}
\rightarrow \textnormal{HCO}^+ + \textnormal{HD}$ unchanged.}.  
The DCO$^{+}$ profile
remains almost the same in the inner 60~AU whether the H$_{2}$D$^{+}$ channel
(reaction~\ref{eq2}) is involved or not. Further out, reaction~\ref{eq2}
contributes for about 40~$\%$ in the DCO$^{+}$ abundance, the other main
contribution being through HCO$^{+}$ (via the reaction
$\textnormal{HC}\textnormal{O}^{+} + \textnormal{D} \rightarrow
\textnormal{DCO}^{+} + \textnormal{H}$, Aikawa et al. in prep.).

\begin{figure*}
\centering
\includegraphics[angle=0,width=8.2cm]{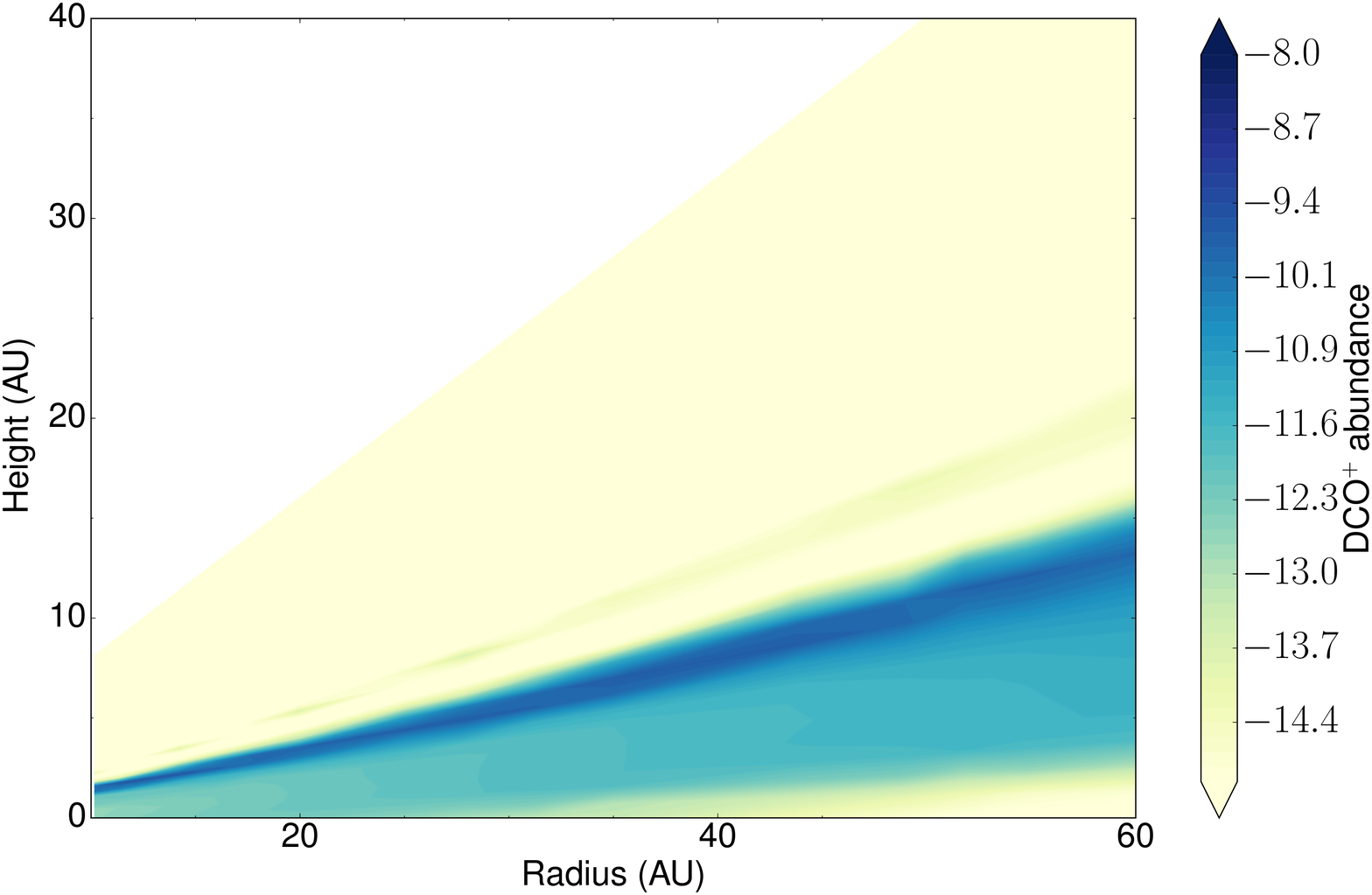}
\includegraphics[angle=0,width=8.2cm]{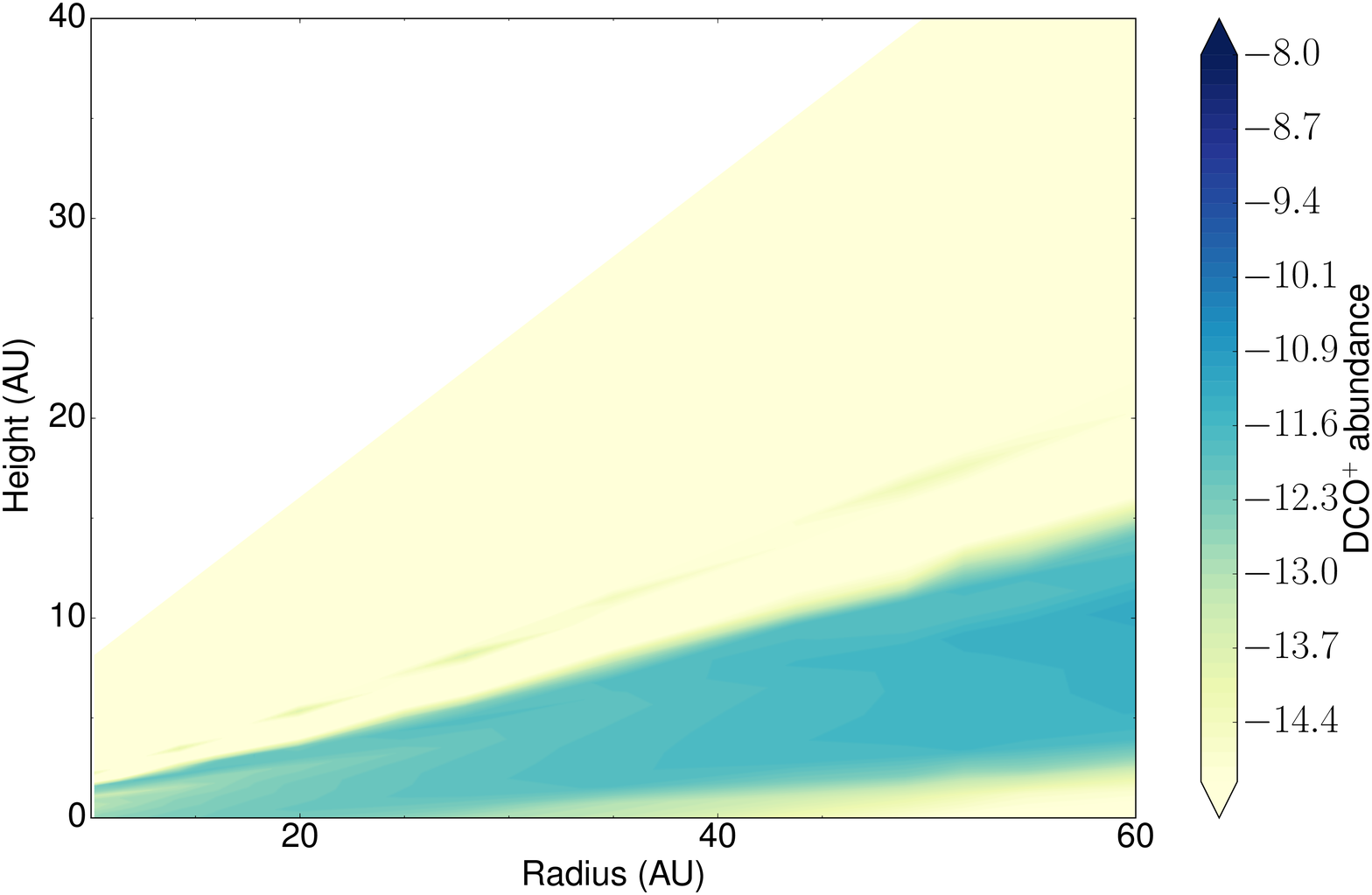}
\includegraphics[angle=0,width=8.2cm]{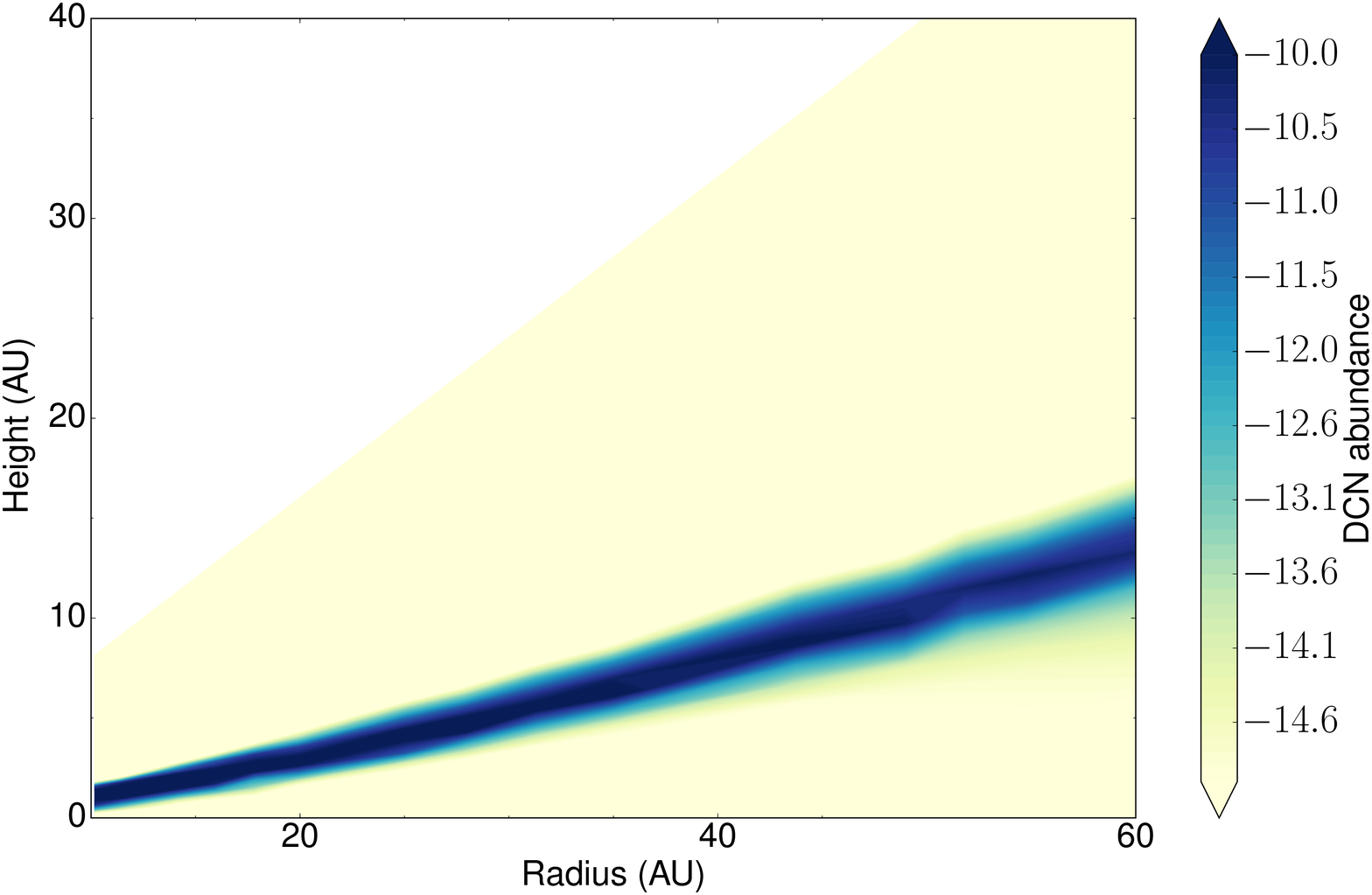}
\includegraphics[angle=0,width=8.2cm]{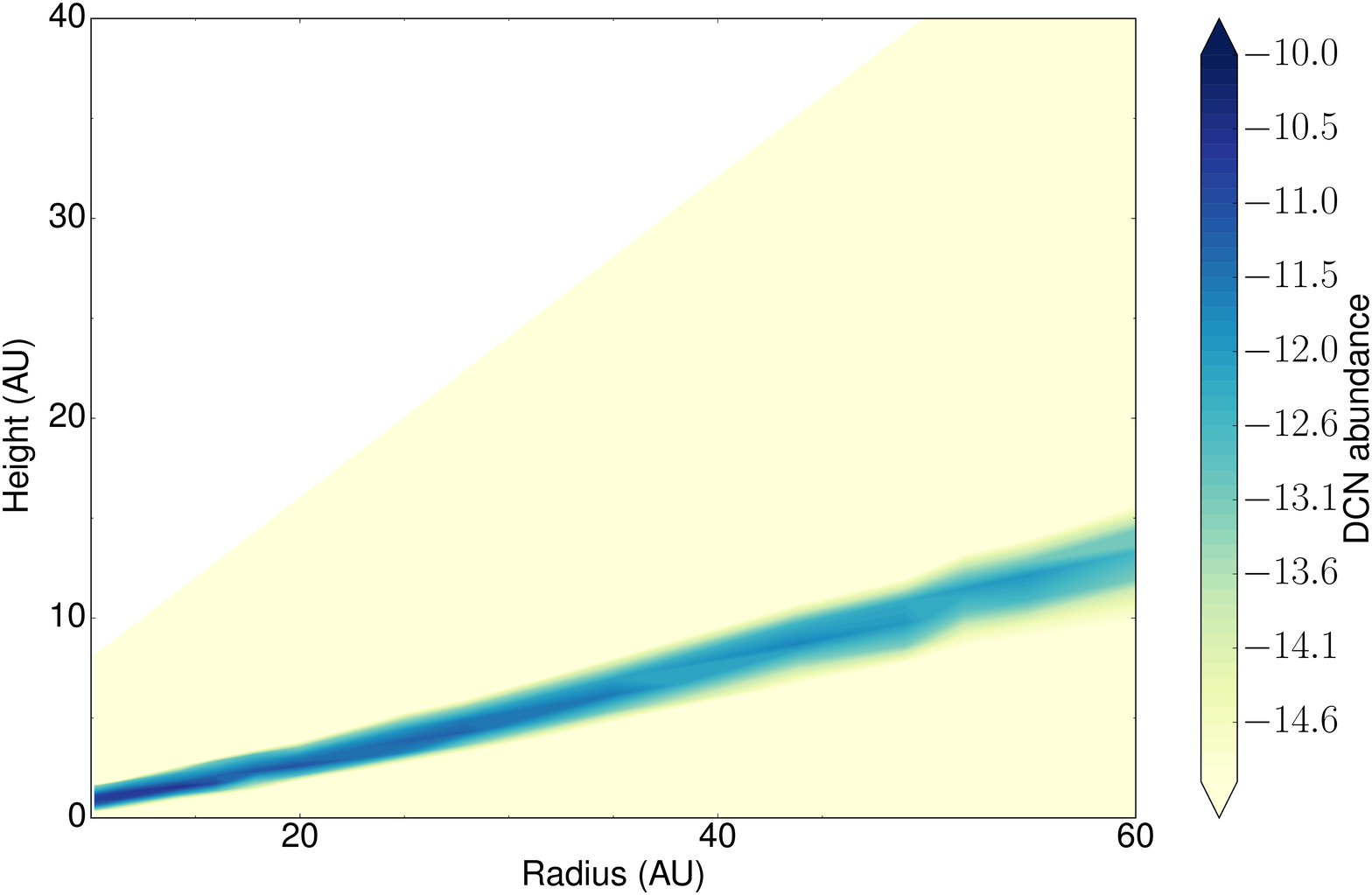}
\includegraphics[angle=0,width=8.2cm]{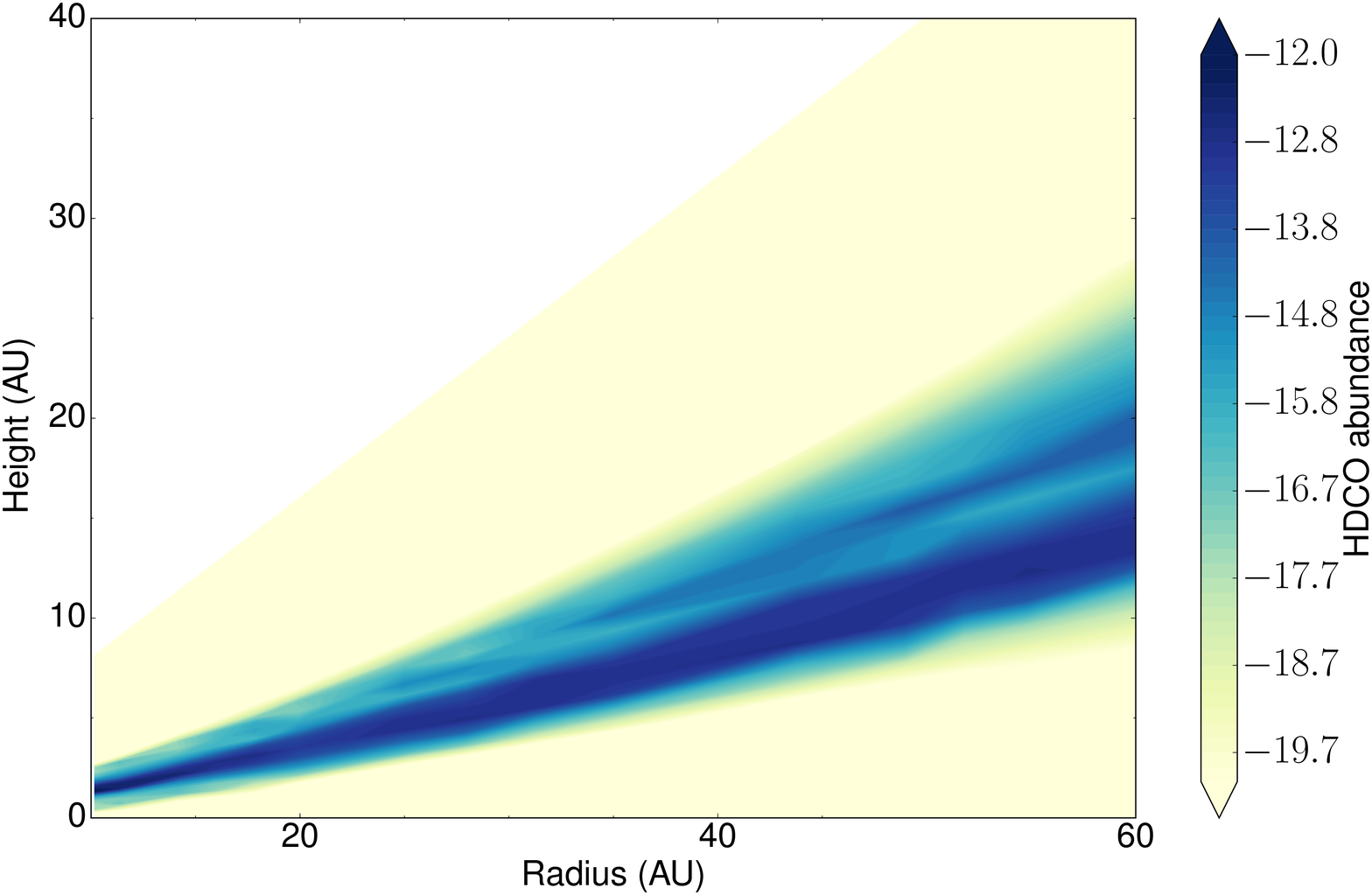}
\includegraphics[angle=0,width=8.2cm]{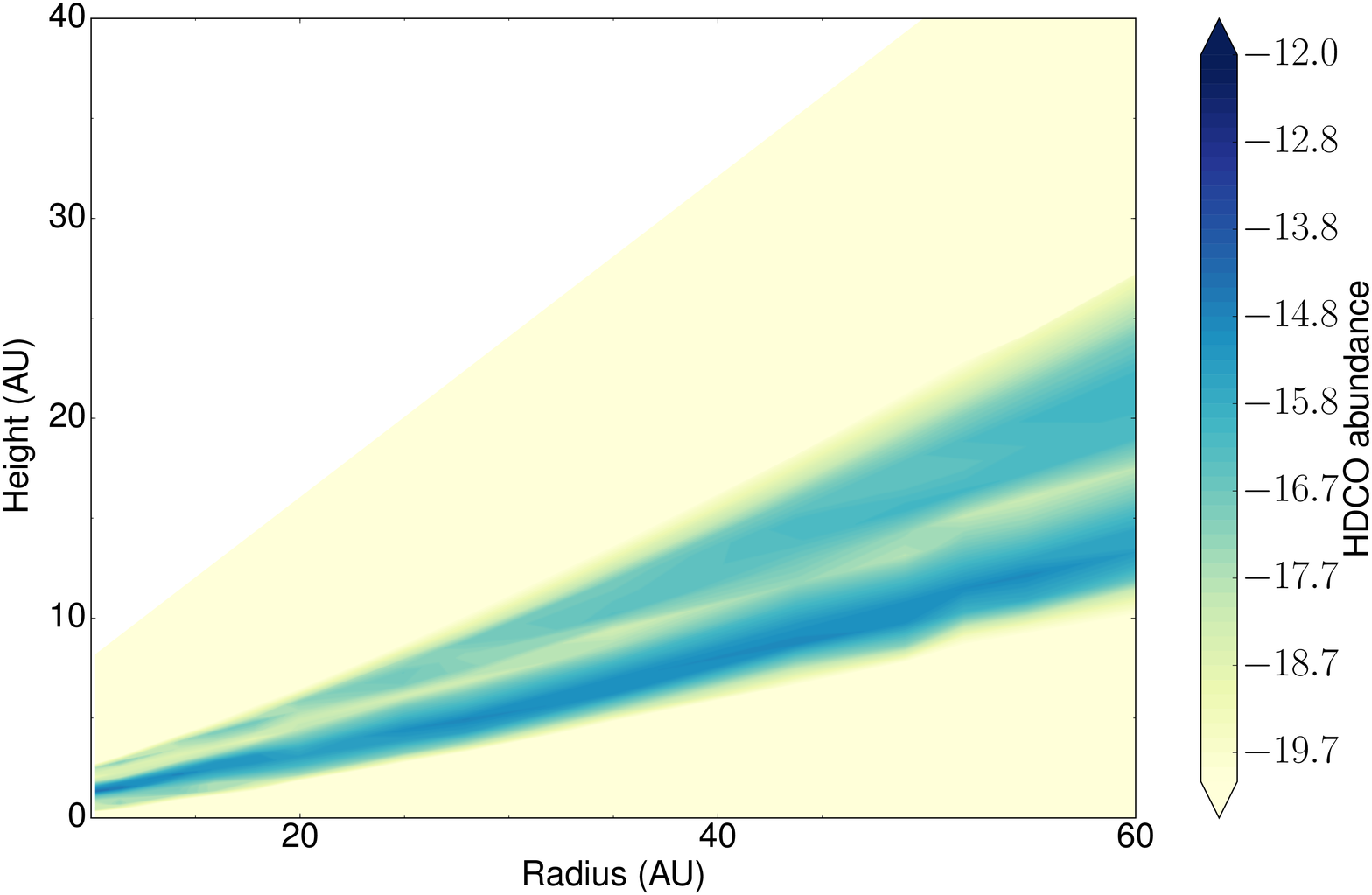}
\caption{ DCO$^{+}$ (top), DCN (middle) and HDCO (bottom) abundance relative to molecular hydrogen. {\em Left panels:} with the CH$_2$D$^{+}$ channel (reaction~\ref{eq3}). {\em Right panels:} without the CH$_2$D$^{+}$ channel.\label{fg3}}
\end{figure*}

\begin{figure*}
\centering
\includegraphics[angle=0,width=12cm]{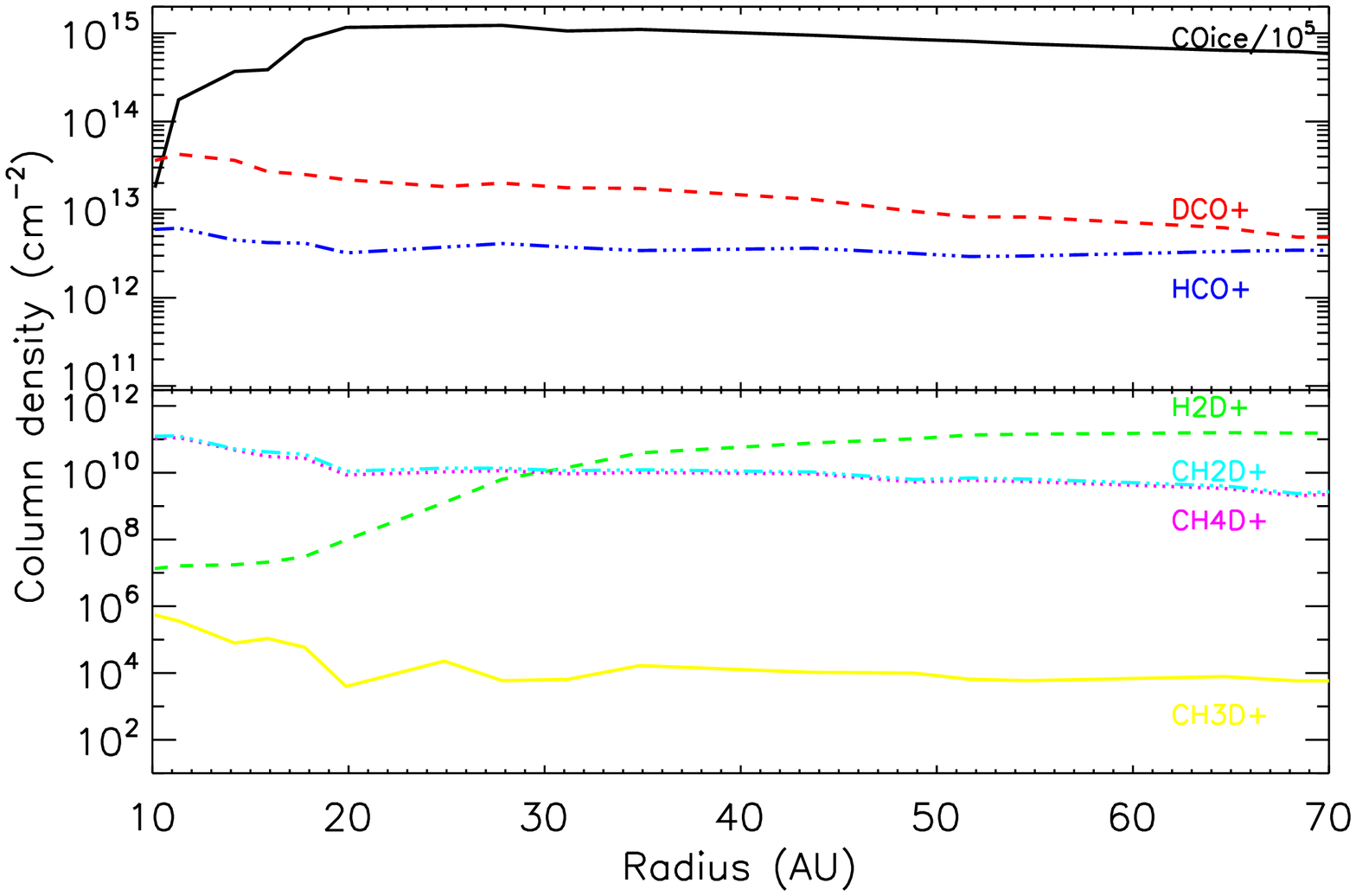}

\includegraphics[angle=0,width=12cm]{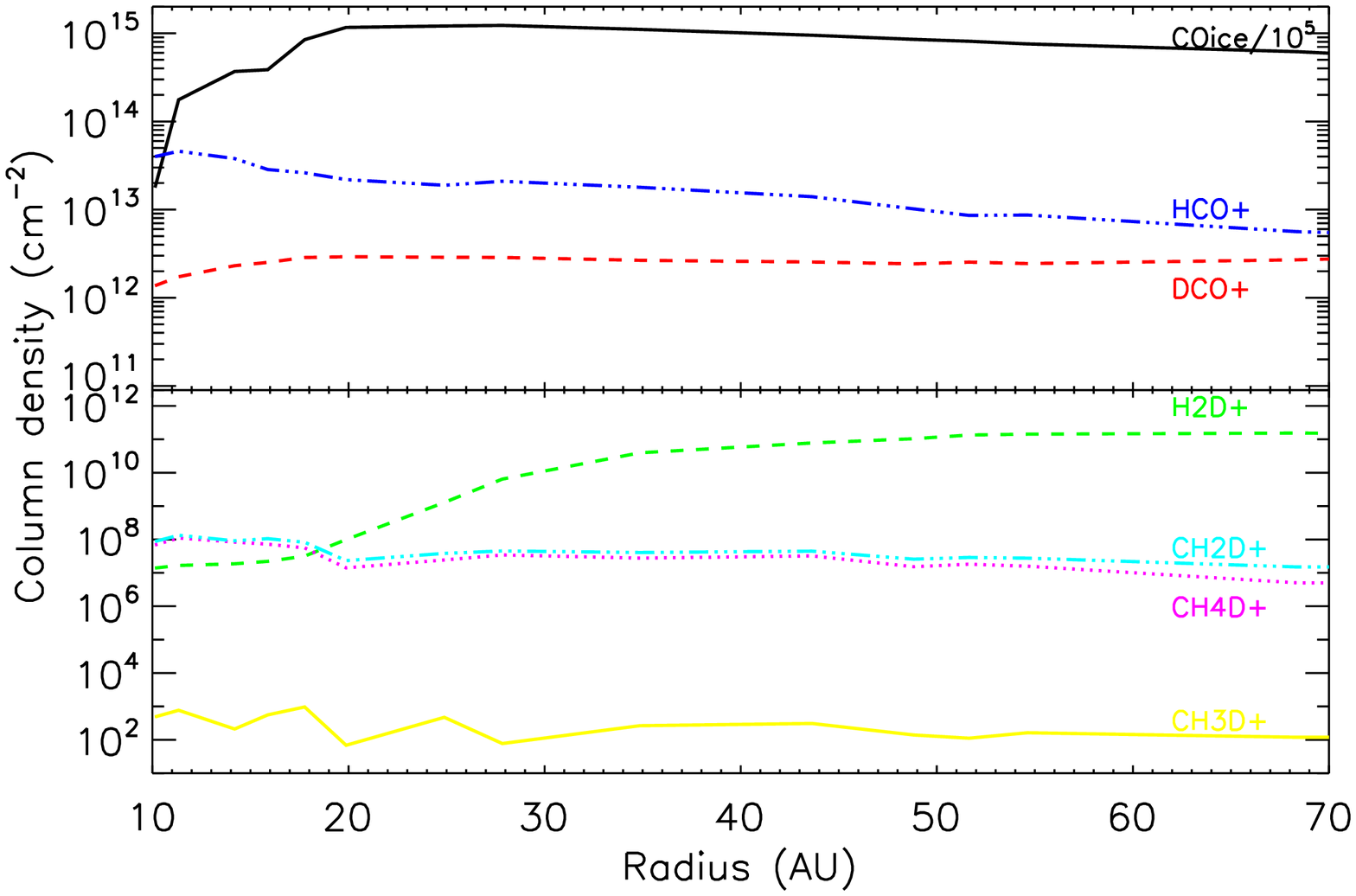}
\caption{Radial distribution of molecular column densities with the CH$_2$D$^{+}$ channel (reaction~\ref{eq3}, top panel) and without the CH$_2$D$^{+}$ channel (bottom panel).{\em Top panel:} The CO$\rm_{ice}$, DCO$^{+}$ and HCO$^{+}$ distribution are indicated in dark solid lines, red dashed lines and blue dash dot dot lines, respectively {\em Bottom panel:} The CH$_{3}$D$^{+}$, H$_{2}$D$^{+}$, CH$_{4}$D$^{+}$ and CH$_{2}$D$^{+}$ are indicated in yellow solid lines, green dashed lines, magenta dotted lines and cyan dash dot dot lines, respectively. Note that in both panels the CO$\rm_{ice}$ column density is divided by a factor 10$^{5}$. \label{fg4}}
\end{figure*}

One notable feature of Figures~\ref{fg3} and~\ref{fg4} is that the formation of
DCO$^{+}$ in the inner surface layers inside of 60~AU involves CH$_2$D$^{+}$ or
one of its product that have previously been created by reaction~\ref{eq3}.
Indeed, Figure~\ref{fg3} shows the abundance of the DCO$^{+}$ ion for up to the 60~AU when reaction~\ref{eq3} is activated (left panel) and turned off (right panel). It is immediately apparent that reaction~\ref{eq3}, which involves the
CH$_2$D$^{+}$ channel, leads to the observed maximal DCO$^{+}$ abundance in the
inner surface layers of the disk. Likewise, a similar result is observed on the abundance profiles of DCN and HDCO (see Figure~\ref{fg3}), molecules that can both also be formed through reaction \ref{eq3}. In addition, in Figure~\ref{fg4}, we give the
radial distribution of some key ions (such as CH$_2$D$^{+}$ and CH$_4$D$^{+}$)
that can lead to the formation of DCO$^{+}$ along with the one for
H$_{2}$D$^{+}$ when reaction~\ref{eq3} is activated (top panel) and turned off
(bottom panel). A decrease in the DCO$^{+}$, CH$_2$D$^{+}$, CH$_3$D$^{+}$ and
CH$_4$D$^{+}$ column density is clearly observed when reaction~\ref{eq3} is not
activated in our network. Finally, we find that more than 80$\%$ of the
DCO$^{+}$ abundance in the inner 60~AU surface layers is due to a DCO$^{+}$
formation through the reactions~\ref{eq3}, \ref{eq4} and \ref{eq5} pathway.

\subsection{DCO$^{+}$ abundance and exothermicity}
\begin{figure*}
\centering
\includegraphics[angle=0,width=13cm]{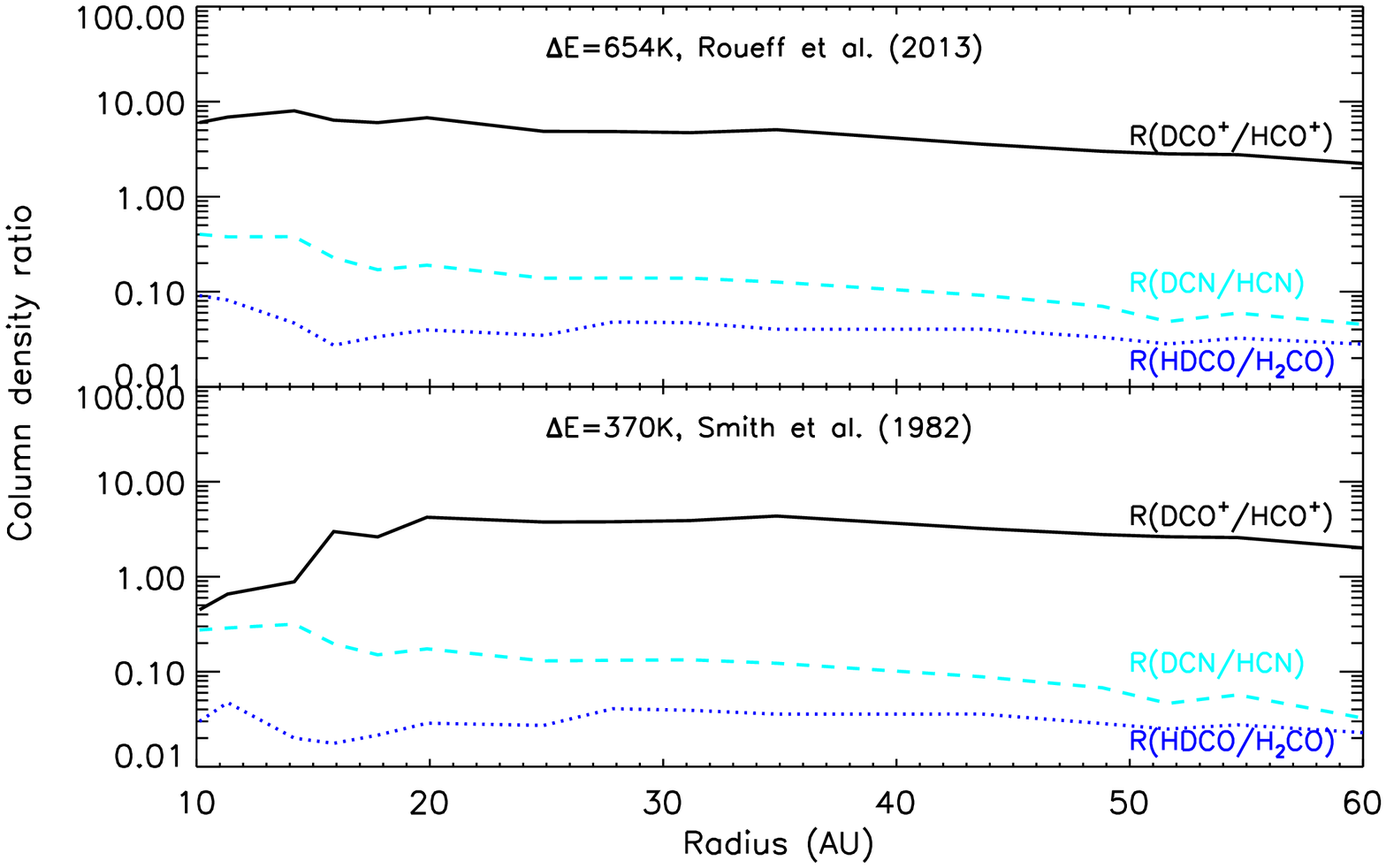}

\caption{Impact on the exothermicity on the DCO$^{+}$/HCO$^{+}$ (dark line), DCN/HCN (cyan dashed line) and HDCO/H$_{2}$CO (blue dotted line) column density ratios. {\em Top panel:} Abundance ratio measured from modeling that includes the new exothermocity of reaction~\ref{eq3} ($\Delta$E = 654~K) set by \citet{Roueff:2013}. {\em Bottom panel:} Abundance ratio measured from modeling that includes the previous exothermicity \citep[$\Delta$E = 370~K, see][]{Smith:1982}.\label{fg5}}
\end{figure*}

In Figure~\ref{fg5}, we show that the use of the new exothermicities leads the warm channel (i.e. reaction \ref{eq3}) to dominate in the DCO$^{+}$ production. More specifically, Fig.~\ref{fg5} illustrates the impact, within the inner disk, of the change of exothermicity on the DCO$^{+}$/HCO$^{+}$ abundance ratio but also on that of other easily observable molecules (DCN, formaldehyde). One notable feature is that the abundance of DCO$^{+}$, DCN and HDCO is enhanced in the 20~AU inner radii when the chemical modeling includes the new exothermicities set by \citet[][$\Delta$E of 654~K for reaction \ref{eq3}]{Roueff:2013} as opposed to the chemical modeling that includes the previous ones \citep[$\Delta$E = 370~K, see][]{Smith:1982}. Furthermore, it is immediately apparent that within the 20~AU inner radii the distribution of the DCO$^{+}$/HCO$^{+}$ abundance ratio is affected by the change of exothermicity: the latter does not increase with increasing radius while using the most recent exothermicities \citep{Roueff:2013}, whereas assuming a lower exothermicity \citep[i.e. the one of][]{Smith:1982} the ratio shows a rise with increasing radius in the inner disk. 

%
\section{Discussion}
\label{sec:conc}

\subsection{The effect of the exothermicity in the CH$_2$D$^{+}$ formation}

Our results are a consequence of the increase in the exothermicity $\Delta$E of
reaction~\ref{eq3}, as suggested by \citet{Roueff:2013}, which favors deuterium
fractionation in warm gas, up to temperatures  even as high as  T $\le$ 300~K. We note that in the case of DCO$^{+}$ only, which has its main formation route from channel \ref{eq5} and, for an exothermicity of 654~K,  a temperature $\ge$71~K  is hot enough to turn off the DCO$^{+}$ enrichment toward the inner disk. Indeed, the activation of the backward
reaction is shifted to higher temperatures, enabling efficient deuteration
closer to the star. This is a nice illustration of the influence of the energy
budget of deuteration reactions. This is similar to the well known effect of
the H$_2$ ortho/para ratio on deuterium chemistry \citep[see
e.g.][]{Flower:2006}: at low temperature, the backward counterpart of
reaction~\ref{eq1} is strongly favored for ortho-H$_2$ over para-H$_2$, due to
a lower $\Delta E$ for the former.  However, we did not consider the spin
states of H$_2$ in the present study. DCO$^+$ being mainly produced by
reaction~\ref{eq3} , we do not expect this omission to be of crucial importance
in the present case (see Sect. 4.4 for further details).


\subsection{DCO$^{+}$ a probe of ionization rather than the CO snow line}
DCO$^{+}$ has been proposed to be a good tracer of the CO snow line by
\citet{Mathews:2013}, assuming H$_2$D$^{+}$ as its parent molecule.  However,
our study points out that the reaction which involves H$_2$D$^{+}$ as a parent
molecule (see reaction~\ref{eq2}) is unlikely to be the main formation route that
leads to the bulk of the DCO$^{+}$ abundance within the disk. Alternatively, our
study shows that reactions involving CH$_2$D$^{+}$ as a parent molecule of
DCO$^{+}$  (see reaction~\ref{eq3} and reactions  \ref{eq4} and \ref{eq5} as
possible pathways for the formation of DCO$^{+}$) are responsible for the
observed high abundance of DCO$^{+}$  in the inner surface layers of the disk.
This finding leads us to argue that, instead of temperature, DCO$^{+}$ is
rather tracing ionization. Indeed, the distribution of the DCO$^{+}$ abundance
is similar to the one of CO except for its high abundance in the the inner
warmer surface layers of the disk where ionization through X-rays is present.
In that respect, we predict that DCO$^+$ is a tracer of the inner tens of AU of
protoplanetary disks surrounding X-ray active T Tauri stars.

\subsection{Implications: on the production of other deuterated molecules}
Chemical pathways involving the CH$_2$D$^{+}$ channel (reaction \ref{eq3}) can sufficiently lead to the formation of different deuterated molecules, such as DCO$^{+}$, DCN, HDCO in warm conditions \citep[e.g. T$\sim$70--100~K, see][]{Wootten:1987,Oberg:2012}. In that context, our study not only shows that the high abundance of DCO$^{+}$ in the inner disk results from pathways involving the CH$_2$D$^{+}$ ion as a parent molecule, but in addition shows that the DCN and HDCO abundance patterns are also affected by the change of exothermicity (see Figs~\ref{fg3} and \ref{fg5}). Indeed, the DCN and HDCO abundances are enhanced while using the CH$_2$D$^{+}$ formation route with the most recent exothermicities \citep[$\Delta$E=654~K,][]{Roueff:2013}. 

\subsection{Implications: on the specific structure of protoplanetary disks}
Interestingly enough, from their ALMA observations, \citet{Mathews:2013} analyzed the DCO$^+$ emission in the disk surrounding HD~163296 as coming from a narrow (confined between 110 and 160 AU) ring-like structure. Although the radial extent of the ring may be subject to revision (Qi et al., in prep), such a narrow ring, located far from the star, is difficult to reconcile with our modeling.
Likewise, based on SMA observations \citep{Qi:2008} of the TW~Hydrae disk and empirical fitting of the data, \citet{Oberg:2012} find a central depression and inferred a DCO$^{+}$ abundance profile that increases with abundance with the radius. These suggestions disagree with our results, since we find that with a higher activation barriers \citep{Roueff:2013} the warm channel (reaction~\ref{eq3}) dominates, leading to a strong abundance of DCO$^{+}$ in the inner regions of the disk.

One possible caveat could be the ortho:para ratio. \citet{Teague:2015} have investigated the effect of the ortho:para ratio on the deuterium fractionation and have found that, by a million years, the o:p ratio is low. However, assuming a higher o:p ratio might change our results by reducing the deuterium enhancement. 
\citet{Roueff:2013} show that the exothermicity in reaction \ref{eq3} is  reduced if there is significant amount of ortho-H$_2$ present in the gas \citep[see Table 3 of][]{Roueff:2013}. As a result, the backward reaction is actually comparable to the old warm channel rates \citep{Smith:1982}. Thus, one way to chemically mimic a hole in the DCO$^{+}$ distribution is to have a gradient in the o:p ratio in the disk which is higher in the inner regions and lower in the outer disk. 

Another possible caveat could be the ionization structure. The presence of structure in the inner disk could perhaps prevent the propagation of the X--ray photons in the inner disk and may lead to a reduced production of DCO$^{+}$. However, this effect may not be as important for the more substantially flared outer disk.

 Finally, the disk surrounding HD~163296 is much warmer than our TW~Hydrae disk model. In this disk, the CO snow line is located around 150~AU \citep{Qi:2011}, while it is around 20~AU in our model. Another possibility would thus be that the HD~163296 disk is warm enough (T$\ge$71~K) and/or present a higher o:p ratio to enable endothermic reactions to prevent deuterium fractionation in the inner 100~AU, thereby truncating the DCO$^+$ layer. Direct modeling of the DCO$^+$ chemistry in the warm disk surrounding HD~163296 would be necessary for further comparisons. 
In any case, further observations of HD~163296  and TW~Hya are required to understand the DCO$^+$ distribution, at the light of our new DCO$^+$ formation pathway.

\section{Conclusions}
To summarize, our study shows that the high abundance of DCO$^{+}$ in the inner surface layers of the disk results from CH$_2$D$^{+}$ as a parent
molecule. The latter being efficiently formed through reaction~\ref{eq3} in
warm conditions (typically T$\le$71~K). As a consequence we suggest that DCO$^{+}$ is a reliable
tracer of ionization in the inner warm disk surface. Furthermore, the change in exothermicity, in the instance of a low o/p ratio, leads to enhancements of other tracers including DCN and HDCO.

%

\acknowledgments
This work was supported by the National Science Foundation under grant 1008800. 
FH thanks the Department of Astronomy (University of Michigan) for its hospitality. 


%
\bibliographystyle{apj}

%

\clearpage

\clearpage

\clearpage

\clearpage

\end{document}